\documentclass[pdflatex,sn-mathphys,iicol]{sn-jnl}% Default with double column layout
\bibstyle{spphys} 

\usepackage{hyperref}
\usepackage{graphicx}
\usepackage[utf8]{inputenc}
\usepackage{amsmath}

\usepackage{physics}

\jyear{2022}

\begin{document}

\title[Scale-Invariant Type II Seesaw Model]{Vacuum Stability and Radiative Symmetry Breaking of the Scale-Invariant Singlet Extension of Type II Seesaw Model}

\author[1]{\fnm{Bayu} \sur{Dirgantara}}\email{bayuquarkquantum@yahoo.com}

\author[2]{\fnm{Kristjan} \sur{Kannike}}\email{kristjan.kannike@cern.ch}

\author*[1]{\fnm{Warintorn} \sur{Sreethawong}}\email{warintorn.s@g.sut.ac.th}

\affil*[1]{\orgdiv{School of Physics and Center of Excellence in High Energy Physics \textit{\&} Astrophysics}, \orgname{Suranaree University of Technology}, \orgaddress{\city{Nakhon Ratchasima}, \postcode{30000}, \country{Thailand}}}

\affil[2]{\orgdiv{Laboratory of High Energy and Computational Physics}, \orgname{NICPB}, \orgaddress{\street{R\"{a}vala}, \city{Tallinn}, \postcode{10143}, \country{Estonia}}}

\abstract{	The questions of the origin of electroweak symmetry breaking and neutrino mass are two major puzzles in particle physics. Neutrino mass generation requires new physics beyond the Standard Model and also suggests reconsideration of physics of symmetry breaking. The aim of this paper is to study radiative symmetry breaking in the singlet scalar extension of type II seesaw neutrino mass model. We derive bounded-from-below conditions for the scalar potential of the model in full generality for the first time. The Gildener-Weinberg approach is utilised in minimising the multiscalar potential. Upon imposing the bounded-from-below and perturbativity conditions, as well as experimental constraints from colliders, we find the parameter space of scalar quartic couplings that can radiatively realise electroweak symmetry breaking at one-loop level. To satisfy all the constraints, the masses of the heavy triplet-like Higgs bosons must be nearly degenerate. The evolution of the Higgs doublet quartic coupling $\lambda_{H}$ can be prevented from being negative up to the Planck scale.}




\maketitle

%=======================================================
\section{Introduction}

The Standard Model (SM) has achieved astounding success in describing fundamental interactions of particles, and its predictions have been persistently tested to high precision. The discovery of a Higgs boson with mass $m_h\simeq$ 125 GeV at the Large Hadron Collider (LHC) \cite{Aad:2012tfa,Chatrchyan:2012ufa} seems to provide the last missing piece of the SM. Nevertheless, unanswered puzzles such as the origin of electroweak symmetry breaking (EWSB), the stability of the Higgs mass scale, the existence of dark matter, and nonzero neutrino masses motivate us to seek new physics beyond the SM (BSM).

In the SM, the electroweak symmetry is spontaneously broken due to the presence of a negative mass term in the Higgs potential. This is the only dimensionful parameter in the theory. New physical states that couple to the Higgs boson can occur anywhere between the electroweak and the Planck scale. Their tree-level and loop-level contributions to the Higgs mass would have to cancel to tremendous accuracy to uphold the hierarchy. 
 Various extensions of the SM aiming to unravel this hierarchy problem involve extra dimensions \cite{Arkani-Hamed:1998jmv, Arkani-Hamed:1998sfv, Randall:1999ee, Randall:1999vf}, a new symmetry such as supersymmetry \cite{Martin:1997ns} or cosmological naturalness, e.g. Ref.~\cite{Csaki:2022zbc}.

An attractive class of models addressing the hierarchy problem stems from inspiring guidance proposed by Bardeen \cite{Bardeen:1995kv}. If the Higgs mass parameter in the SM is forbidden by classical scale invariance, which is broken only by quantum anomalies, the hierarchy problem can be alleviated. (Besides that, if we assume that physics at the Planck scale -- quantum gravity -- behaves differently from usual quantum field theory, there should be no intermediate scale between the electroweak scale and the Planck scale, and  no instability or Landau pole before the Planck scale~\cite{Meissner:2007xv}.)  In this way, a mass scale can be dynamically generated in model with classical scale symmetry via dimensional transmutation as first demonstrated in a seminal paper by S. Coleman and E. Weinberg~\cite{Coleman:1973jx}. Unfortunately, the radiative EWSB via the Coleman-Weinberg (CW) mechanism can not be realized in classically scale-invariant SM since the top quark renders the one-loop Higgs potential unbounded from below \cite{Fujikawa:1978ru}. Nevertheless, a plethora of proposals have been putting forward a scale invariance with extended scalar sector as a possible solution to the hierarchy problem \cite{Foot:2007as,Espinosa:2007qk,Foot:2007iy,Iso:2009ss,Foot:2010av,AlexanderNunneley:2010nw,Farzinnia:2013pga,Heikinheimo:2013fta,Karam:2015jta,Ghorbani:2017lyk}. In the case of multi-field potentials, the minimum direction can be found by the Gildener-Weinberg (GW) method \cite{Gildener:1976ih}.

On the other hand, the discovery of neutrino oscillations have provided us the solid evidence of massive neutrino \cite{SNO:2001kpb, SNO:2002tuh,K2K:2006yov,KamLAND:2002uet}. One of the appealing BSM extensions that can naturally induce the tiny
neutrino masses is the type II seesaw model \cite{Magg:1980ut,Schechter:1980gr,Cheng:1980qt,Lazarides:1980nt,Mohapatra:1980yp}, in which the Higgs sector is extended by an $SU(2)_L$ Higgs triplet. A trilinear interaction between the doublet and triplet Higgs plays an important role in generating Majorana neutrino masses and is the source of lepton number violation in the model. However if classical scale invariance is imposed, this trilinear term is forbidden and a global lepton number symmetry will be spontaneously broken after the triplet develops nonzero vacuum expectation value (VEV). This results in the emergence of a massless Goldstone boson, a majoron \cite{Chikashige:1980ui}. Since a triplet majoron has $SU(2)_L$ and $U_Y(1)$ gauge interactions, it affects the invisible decay width of the $Z$ boson and has already been ruled out \cite{GonzalezGarcia:1989zh}. A majoron that arises predominantly from a singlet \cite{Masiero:1990uj}, however, is still allowed.

In this work, we consider a scalar singlet extension of the type II seesaw model with a classical scale-invariant scalar potential. This model was originally proposed in  Ref.~\cite{Schechter:1981cv} without classical scale symmetry, and its collider phenomenology was studied in~\cite{Diaz:1998zg,Bonilla:2015jdf}. With the aid of the orbit space of scalar quartic gauge invariants -- in particular the $P$-matrix method \cite{Talamini:2006wd,Abud:1983id,Abud:1981tf} -- we derive vacuum stability constraints and study the radiative EWSB along the flat direction of the tree-level scalar potential. We find a range of VEVs and particle masses that realises the EWSB and is compatible with all theoretical and experimental constraints.

This paper is organised as follows. In Sec.~\ref{sec:model} we briefly review the type II seesaw model and introduce its scale-invariant singlet extension. 
In Sec.~\ref{sec:orbit:space}, we determine the orbit space of the model. In Sec.~\ref{sec:BfB} we study the sufficient and necessary conditions for the scalar potential to be bounded from below with details given in Appendix~\ref{app:BfB2}. In Sec.~\ref{sec:RSB}, the effective potential is minimised via GW method. We show the available parameter space in Sec.~\ref{sec:result} and present our conclusions in Sec.~\ref{sec:conclusions}.

%%%%%%%%%%%%%%%%%%%%%%%%%%%%%%%%%%%%%%
\section{Scale-Invariant Extension of the Type II Seesaw Model}
\label{sec:model}

Considering the SM as an effective field theory, one can add higher-dimensional operators which encode the effect of heavy degrees of freedom in UV-complete theory to low energy physics. The Weinberg operator $LLHH$ is a unique dimension-5 operator that can generate neutrino mass after spontaneous symmetry breaking. The tree-level realisations of this operator are classified into three types of canonical seesaw models~\cite{Ma:1998dn}. Among the seesaw model variants, the type II seesaw model offers a rich phenomenology to study. 
However, it fails to be a scale-invariant model that could address the hierarchy problem. In addition to the SM-Higgs doublet mass term, there are two additional dimensionful parameters entering scalar potential of type II seesaw: the triplet mass term and the trilinear coupling between doublet and triplet fields:
\begin{equation}
\begin{split}
  V &= \mu^{2}_{H} H^{\dagger} H + \mu^{2}_{\Delta} \Tr (\Delta^{\dagger} \Delta)
  + \lambda_{H} (H^{\dagger} H)^{2}  \\
  & + \lambda_{\Delta} \Tr (\Delta^{\dagger} \Delta)^{2} + \lambda'_{\Delta} \Tr (\Delta^{\dagger} \Delta \Delta^{\dagger} \Delta) \\
  &+ \lambda_{H\Delta} H^{\dagger} H \Tr (\Delta^{\dagger} \Delta)
		+ \lambda'_{H\Delta} H^{\dagger} \Delta \Delta^{\dagger} H
		\\
		&	+ \frac{1}{2} (\mu H^{T} \varepsilon \Delta^{\dagger} H + \text{h.c.}),
\end{split}
\label{eq:V:II}
\end{equation}
where $H$ is the SM Higgs doublet with hypercharge $Y = 1$ and lepton number $L = 0$ and $\Delta$ is an $SU(2)$ triplet with hypercharge $Y = 2$ and $L = -2$. Notice that the presence of the trilinear coupling $\mu$ explicitly breaks the lepton number invariance. 

In order to construct a classically scale-invariant model of type II seesaw, we consider, besides $H$ and $\Delta$, a complex singlet $S$ with $L = -2$. Then, the dimensionful terms in the potential can be generated when a scalar singlet $S$ gets a VEV. This model was originally proposed in  Ref.~\cite{Schechter:1981cv} without classical scale symmetry, its collider phenomenology was studied in Refs.~\cite{Diaz:1998zg,Bonilla:2015jdf} and a recent review is given by Ref. \cite{Mandal:2022zmy}.

We parametrise the Higgs fields around the neutral electroweak minimum as
\begin{align}
	S &= \frac{1}{\sqrt{2}}(v_{s}+S_{R}+i S_{I}),
	\\
	H & = \begin{pmatrix} 
	h^{+}
	\\
	\frac{v_{h}+ \phi_{h} +i \chi_{h}}{\sqrt{2}}
	\end{pmatrix},
	\\
	\Delta &\equiv \frac{\vec{\sigma}}{\sqrt{2}}\cdot \vec{\Delta} = \begin{pmatrix}  \delta^{+}/\sqrt{2} & \delta^{++} \\
		\frac{v_{\delta}+ \phi_{\delta} +i \chi_{\delta}}{\sqrt{2}}&  -\delta^{+}/\sqrt{2}
	\end{pmatrix},
\end{align}
where $v_s, v_h$ and $v_{\delta}$ are the VEVs of the singlet, doublet and triplet, respectively, and $\vec{\sigma}$ are the Pauli matrices.

With classical scale invariance, the most general renormalisable scalar potential takes the form 
\begin{equation}\label{eq:vpot}
	\begin{split}
		V &= \lambda_{H} (H^{\dagger} H)^{2} + \lambda_{S} (S^{\dagger} S)^{2} \\
		&+ \lambda_{\Delta} \Tr (\Delta^{\dagger} \Delta)^{2} 
		+ \lambda'_{\Delta} \Tr (\Delta^{\dagger} \Delta \Delta^{\dagger} \Delta)
		\\
		&+ \lambda_{H\Delta} H^{\dagger} H \Tr (\Delta^{\dagger} \Delta)
		+ \lambda'_{H\Delta} H^{\dagger} \Delta \Delta^{\dagger} H
		\\
		&+ \lambda_{HS} H^{\dagger} H S^{\dagger} S + \lambda_{S\Delta} S^{\dagger} S \Tr (\Delta^{\dagger} \Delta) 
		\\
		&+ \frac{1}{2} (\lambda_{SH\Delta} S H^{T} \varepsilon \Delta^{\dagger} H + \text{h.c.}),
	\end{split}
\end{equation}
where all the couplings are real except $\lambda_{SH\Delta}$, which we make real as well by a phase rotation without loss of generality. The scale-invariant potential \eqref{eq:vpot} also respects lepton number. (See ref.~\cite{Okada:2015gia} for the scale-invariant type II seesaw model with the extended gauge group $U(1)_{B-L}$). After $S$ and $\Delta$ develop VEVs, the global lepton number symmetry will be spontaneously broken, resulting in an emergence of massless Goldstone boson -- the majoron. In this case, a majoron is mainly singlet under the SM gauge interactions. All in all, the physical mass eigenstates comprise the charged scalars $H^{\pm\pm} \equiv \delta^{\pm\pm}$ and $H^{\pm}$,  the neutral CP-even scalars $\varphi$, $h$, $H$ and the CP-odd scalars $J$ and $A$. The mass spectrum and mixing matrices are given in Sec.~\ref{sec:spectra}.

%%%%%%%%%%%%%%%%%%%%%%%%%%%%%%%%%%%%%%
\section{Orbit Space}\label{sec:orbit:space}

We now turn our attention to the constraints on scalar quartic couplings required by the vacuum stability of the scalar potential. To ensure a finite minimum, the potential must be bounded from below (BfB) in all possible directions of the field space as the fields become large. In multi-scalar theories, finding vacuum stability conditions or potential minima is a non-trivial task. 

A powerful method to deal with this is to write the scalar potential in terms of gauge invariant variables: the norms of fields (or their ratios) and angular variables known as orbit space parameters ~\cite{Abud:1981tf,Kim:1981xu,Abud:1983id,Kim:1983mc}. The physical region of orbit parameters is called the orbit space. An elegance of this method is that it contains all the information needed to determine the minimum of potential. More interestingly, when potential is monotonous function of orbit space parameters, its minimum is located on the boundary of the orbit space.

\subsection{Orbit Space and Its Boundary}

The components of a constant scalar field configuration $\phi$ (such as a VEV) will rotate amongst themselves under a gauge transformation $T(\theta)$  through a gauge orbit: $\phi \to \phi_{\theta} = T(\theta) \phi$. The value of the scalar potential $V(\phi)$ or any other gauge-invariant function, on the other hand, remains the same. In particular, for a unitary group all the states $\phi_{\theta}$ have the same norm $\phi_{i}^{*} \phi_{i}$. 

For a compact group, all gauge-invariant polynomials constructed of scalar fields can be given as combinations of elements of a finite polynomial basis (minimal integrity basis) of the orbit space: $p_{a}$ with $a = 1, \ldots, q$. In particular, we can write the scalar potential in terms of this basis, whose elements comprise a finite number of gauge invariants including the norms of fields.
 Because the basis does not change under gauge transformations, a gauge orbit corresponds to a single point in the orbit space.

The orbit space of a compact group is a closed connected subset of $\mathbb{R}^{q}$ with $q$ the number of the polynomials in the minimal basis. It can be described by a finite number of polynomial equations or inequalities. It is useful to reduce the orbit space to unit norms of fields by defining dimensionless ratios of the or orbit space variables such as 
\begin{align}
	\alpha &= \frac{f_{ijkl}\phi_{i}^{*}\phi_{j}\phi_{k}^{*}\phi_{l}}{(\phi_{m}^{*}\phi_{m})^{2}},
\end{align}
where $f^a_{ijkl}$ denotes a gauge contraction \cite{ElKaffas:2006gdt,Arhrib:2011uy,Bonilla:2015eha}. In this way, we can write the scalar potential in terms of field norms $\phi_{m}^{*}\phi_{m}$ and the orbit space variables. Below, it will be clear from the context whether we mean by the orbit space the space of the basis polynomials or the reduced space of the dimensionless orbit variables.

Each subgroup of the full gauge group $G$ is the isotropy subgroup $G_{\phi}$ of some field configuration $\phi$. Moreover, all the transformed states $\phi_{\theta}$ have the same isotropy subgroup $G_{\phi}$. The set of orbits that respects the same isotropy subgroup is called the stratum of the isotropy subgroup. The VEV $\phi$ of the potential that breaks the full gauge group $G$ to $G_{\phi}$ therefore lies in the stratum of $G_{\phi}$. In the main stratum -- corresponding to a general field configuration -- the gauge symmetry is completely broken, while the lower-dimensional strata that form the orbit space boundary correspond to more symmetrical field configurations invariant under larger isotropy subgroups $G_{\phi}$. The orbit space thus consists of strata of different dimensions: vertices, edges, \ldots, up to the main stratum whose dimension is given by the number of orbit space variables.  For three orbit space variables, as in our case, the main stratum is three-dimensional and the boundary of the orbit space has two-dimensional faces bordered by edges which end at the vertices of the orbit space.

We derive the boundary of the orbit space using two methods. First of all, in a conventional approach, the set of equations describing the boundary of orbit space can be obtained  by trial and error by taking particular field components to zero. 

A more powerful approach is the so-called $P$-matrix method  \cite{Talamini:2006wd,Abud:1983id,Abud:1981tf}. The $P$-matrix is a $q \times q$ symmetric and positive semi-definite matrix with elements constructed from gradients of basis invariants $p_{a}$, given by
\begin{equation}
	P_{ab} = \frac{\partial p_{a}}{\partial \phi_{i}^{\dagger}} \frac{\partial p_{b}}{\partial \phi_{i}},\label{pmatrix}
\end{equation}
where $\phi_{i}$ run over the field components. Essentially it is the Hermitian square of the Jacobian matrix. It can be shown that elements of the $P$-matrix can be given in terms of the minimal integrity basis $p_{a}$.

The $P$-matrix is positive-definite only inside the orbit space. For that reason, the boundary of the orbit space is obtained by solving $\det P=0$, which is a polynomial equation in the basis elements $p_{a}$. In particular, the orbit space vertices are found by requiring that all the one-by-one principal minors of the $P$-matrix vanish; the edges, by requiring that the two-by-two principal minors vanish (with the one-by-one principal minors positive); etc. When the orbit space has more than three dimensions, then the $P$-matrix approach is much more efficient.

We hope that this necessarily very cursory overview of the orbit space may be enough for an intuitive understanding of the next subsections and refer the interested reader for details to the original references \cite{Talamini:2006wd,Abud:1983id,Abud:1981tf}. 

%%%%%%%%%%%%%%%%%%%%%%%%%%%%%%%%%%%%%%
\subsection{Orbit Space Parameters}

Through the gauge invariants present in the potential \eqref{eq:vpot}, we define the orbit space parameters $s,h, \delta,\zeta,\xi,\eta,\alpha$ as follows\footnote{In Sec. \ref{sec:RSB}, we denote by $h$ the usual physical Higgs boson, as will be clear from the context.}
\begin{align}
	H^{\dagger}H &\equiv h^2 \label{eq2},\\ 
	S^{\dagger}S &\equiv s^2\label{eq3},\\
	\Tr(\Delta^{\dagger}\Delta)^{2} &\equiv \delta^2,
	\\
	(\Tr\Delta^{\dagger}\Delta)^{2} &\equiv  \zeta \Tr(\Delta^{\dagger}\Delta)^{2} \; \label{eq4},\\ 
	H^{\dagger}\Delta\Delta^{\dagger}H &\equiv  \xi \; (H^{\dagger}H)\Tr(\Delta^{\dagger}\Delta)\label{eq5},\\ 
	S H^{T}\epsilon \Delta^{\dagger}H &\equiv  \eta e^{i\alpha} \, H^{\dagger}H \sqrt{S^{\dagger}S} \sqrt{\Tr(\Delta^{\dagger}\Delta)}.
	\label{eq6}
\end{align}
By considering simplest field configurations, with most of the field components set to zero, the ranges of these orbit parameters are found to be
\begin{equation}
  \begin{aligned}
	0 &\leq h, 
	&
	0 &\leq s, 
	&
	0 &\leq \delta,
	\\
	1/2 \leq \zeta &\leq 1,
	&
	0\leq \xi &\leq 1, 
	&
	0 \leq \eta &\leq 1,	
	\\
	0\leq \alpha & < 2\pi.
\end{aligned}
\end{equation}

In terms of orbit space parameters, the potential \eqref{eq:vpot} reads
\begin{equation}
\label{eq:V:orbit}
\begin{split}
	V &= \lambda_{H} h^{4} + \lambda_{S} s^{4}
	+ (\lambda_{\Delta} + \lambda'_{\Delta} \zeta) \delta^{4}  
	\\
	&+ (\lambda_{H\Delta} + \lambda'_{H\Delta} \xi) h^{2} \delta^{2}
	+ \lambda_{HS} h^{2} s^{2}  \\
	&+ \lambda_{S\Delta} s^{2} \delta^{2} + \abs{\lambda_{SH\Delta}} \eta s \delta h^{2} \cos \alpha.
\end{split}
\end{equation}

Because the potential \eqref{eq:V:orbit} is linear in $\xi$, $\zeta$ and $\eta$, the potential minimum is on the boundary of the orbit space -- more precisely, on the intersection of the orbit space with its convex hull \cite{Kim:1981xu,Degee:2012sk,Heikinheimo:2017nth}. Note that one does not have to separately minimise the potential over any flat or concave regions of the orbit space, since such a region is already accounted for in the convex hull by its edges. For shortness, we will denote a vector of the three orbit space parameters as $\vec{\rho}=(\xi,\zeta,\eta)$. The last term of the potential \eqref{eq:V:orbit} satisfies
\begin{equation}
	\min \abs{\lambda_{SH\Delta}} \eta s \delta h^{2} \cos \alpha
	= -\abs{\lambda_{SH\Delta}} \eta s \delta h^{2}
	\label{eq:alpha:min}
\end{equation}
in the potential minimum, so the three parameters in $\vec{\rho}$ suffice.

In the conventional approach, we obtain four non-trivial boundary solutions by taking all possible pairs of fields to be zero. As an example, if one consider the direction where $\delta^{+}$ and $h^{+}$ vanish, one gets
\begin{align}
	\lim_{\delta^{+},h^{+}\to 0}\eta&= \sqrt{\xi},\label{eq27}\\
	\lim_{\delta^{+},h^{+}\to 0}\zeta&= 2\eta^{4}-2\eta^{2}+1.\label{eq28}
\end{align}%27-28% 
The first boundary solution is then expressed in parametric form as 
\begin{equation}
	\vec{\rho}_{\rm I}=(\xi,2\xi^2-2\xi+1,\sqrt{\xi}), \quad 0\le\xi\le 1.
	\label{eq:edge:I}
\end{equation}
The curve $\vec{\rho}_{\rm I}$ is an edge of the orbit space. 
The remaining three edges can be obtained in similar fashion: 
\begin{align}
  \vec{\rho}_{\rm II} &= (\xi,1-2\xi^2,0), & 0\le\xi\le 1/2, 
  \label{eq:edge:II}
  \\
  \vec{\rho}_{\rm III} &= (\xi,1,\xi), & 0\le\xi\le 1,
  \label{eq:edge:III}
  \\
  \vec{\rho}_{\rm IV} &= (1/2,1/2,\eta), & 0\le\eta\le 1/\sqrt{2}.
  \label{eq:edge:IV}
\end{align}

%%%%%%%%%%%%%%%%%%%%%%%%%%%%%%%%%%%%%%
\subsection{$P$-matrix Approach}

We will now determine the whole orbit space via the $P$-matrix approach.
We define gauge invariant polynomials $p_1$ to $p_6$ that enter the scalar potential as 
\begin{align}
	p_{1} &= S^{\dagger} S \equiv s^{2},
	\\
	p_{2} &= H^{\dagger} H \equiv h^{2},
	\\
	p_{3} &= \tr (\Delta^{\dagger} \Delta) \equiv \delta^{2},
	\\
	p_{4} &= H^{\dagger} \Delta \Delta^{\dagger} H \equiv \xi h^{2} \delta^{2},
	\\
	p_{5} &= \tr (\Delta^{\dagger} \Delta \Delta^{\dagger} \Delta) \equiv \zeta \delta^{4},
	\\
	p_{6R} + i p_{6I} &= S H^{T} \epsilon \Delta^{\dagger} H \equiv \eta e^{i \alpha} s \delta h^{2},
\end{align}
where the parameters $\xi$, $\zeta$, $\eta$ and $\alpha$ are the same as in Eqs.~\eqref{eq4}-\eqref{eq6}. 
Thanks to Eq. \eqref{eq:alpha:min}, we can consider the absolute value of $p_6$
\begin{equation}
\begin{split}
	\abs{p_{6}}^{2} &= p_{6R}^{2} + p_{6I}^{2} = \abs{S H^{T} \varepsilon \Delta^{\dagger} H}^{2}
	\\
	& = \eta^{2} s^{2} \delta^{2} h^{4}
\end{split}
\end{equation}
instead of separate $p_{6R}$ and  $p_{6I}$.
We calculate the elements of the $P$-matrix defined in Eq.~(\ref{pmatrix}) where $p_{a}$ are given by $p_{1}$ to $p_{5}$ and $\abs{p_{6}}^{2}$.
In general, the $P$-matrix elements are gauge-invariant quantities, and can be expressed in terms of the gauge invariant polynomials. For the present model, unfortunately, our polynomial basis is not complete. To complete the basis would necessitate introducing higher-order ($d > 4$) invariants which would complicate things considerably. However, we can find an equation for the boundary of the orbit space directly in terms of field components. In this approach, we express the $SU(2)$ triplet as a complex traceless matrix of the form $\Delta = \frac{\vec{\sigma}}{\sqrt{2}}\cdot\vec{\Delta}$. We can use an $SU(2)$ gauge rotation to get rid of three real components of the triplet, and parametrise the remaining components as
\begin{equation}
	\Delta_{1} = x, \quad \Delta_{2} = i y, \quad \Delta_{3} = z,
\end{equation}
so that the norm of $\Delta$ is given by $\delta^{2} = x^{2} + y^{2} + z^{2}$.

It is easy to show that the orbit space parameters can in principle only depend on the difference of the phases of the two components $h_{1}$ and $h_{2}$  of the Higgs doublet. Real solutions for real components of the fields, however, are only obtained when the phase difference is zero. For that reason, we take $h_{1}$ and $h_{2}$ to be real on the orbit space boundary without loss of generality. The orbit space parameters on the orbit space boundary are given by
\begin{align}
  \xi &= \frac{1}{2} + \frac{y (h_{1}^2 x - h_{2}^2 x - 2 h_{1} h_{2} z)}{(h_{1}^2 + h_{2}^2) (x^2 + y^2 + z^2)},
  \\
  \zeta &= \frac{1}{2} + \frac{2 y^2 (x^2 + z^2)}{(x^2 + y^2 + z^2)^2},
  \\
  \eta & = \frac{\abs{h_{2}^2 (y - x) + h_{1}^2 (x + y) - 2 h_{1} h_{2} z}}{\sqrt{2} \sqrt{x^2 + y^2 + z^2} (h_{1}^2 + h_{2}^2)}.
\end{align}

The equation $\det P = 0$ for the boundary of the orbit space is then given by
\begin{equation}
\begin{split}
	&y (x^2 - y^2 + z^2) \, (4 x^2 + 4 z^2 + h_{1}^2 + h_{2}^2)
	\\
	&\times [2 x h_{1} h_{2} + z (h_{1}^2 - h_{2}^2)]
	\\
	&\times [(x + y) h_{1}^2 - 2 z h_{1} h_{2} + (y - x) h_{2}^2] = 0.
\end{split}
	\label{eq:orbit:space:boundary}
\end{equation}
The boundary equation \eqref{eq:orbit:space:boundary} has 8 real solutions. Some of them are different parametrisations of the same strata; in the end, four distinct edges are obtained, coinciding with the results \eqref{eq:edge:I}, \eqref{eq:edge:II}, \eqref{eq:edge:III}, and \eqref{eq:edge:IV} obtained from conventional method. 
%In details, edge I is given e.g. by $h_{2} = 0, z = 0$, edge II is given by $z^{2} = x^{2} + y^{2}$, edge III is given by $y = 0$, and edge IV is given by $h_{1} = 0, z = 0$ or $h_{2} = 0, y = -x$.

\begin{figure*}[tb]
	\begin{center}
		\includegraphics{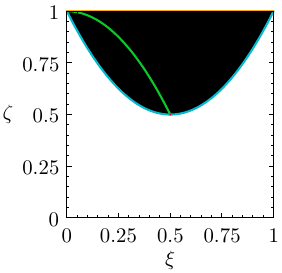}
		\includegraphics{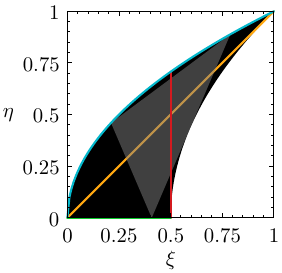}
		\includegraphics{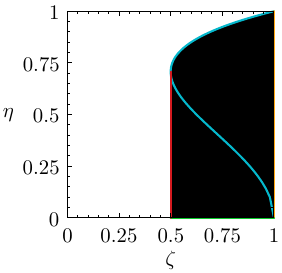}
		\caption{Two-dimensional projections of the orbit space on the $\xi\zeta$-, $\xi\eta$- and $\zeta\eta$-planes, respectively. Boundary solutions I, II, III, and IV are shown in blue, green, yellow and red, respectively. The middle panel also shows the constant $\zeta = \frac{2}{3}$ triangular slice of the orbit space in gray. The projection on the $\xi\eta$-plane is the union of such slices. The vertex A that yields physical EWSB is projected to the upper-right corner of each plot.}
		\label{fig:xi:zeta}
	\end{center}
\end{figure*}

The orbit space has three vertices at the ends of the edges:
\begin{equation}
    \vec{\rho}_{\text{A}} =  (1, 1, 1),
	\;
	\vec{\rho}_{\text{B}} =  ({\scriptstyle \frac{1}{2}}, {\scriptstyle \frac{1}{2}}, 0),
	\;
    \vec{\rho}_{\text{C}} = (0, 1, 0).
    \label{eq:vertices}
\end{equation}
The  $h_{1} = h_{2} \to 0$ limit solution gives the two-dimensional surface of the orbit space. If we take a section of this surface at constant $\zeta$, we obtain a triangle  on the $\xi\eta$-plane whose vertices are given by the intersection points of the $\zeta = \text{const}$ plane with the edge I (in two places) and edge II. The two straight edges III and IV are the degenerate limiting cases of this triangle at extremal values of $\zeta$. The vertices of the triangle at a given $\zeta$ are given by
\begin{align}
	\vec{\rho}_{0} &= \left(
	\frac{1}{\sqrt{2}} \sqrt{1 - \zeta}, \;  \zeta, \; 0 \right),
	\label{eq:II:crossing:zeta}
	\\
	\vec{\rho}_{\pm} &= \left(
	\frac{1 \pm \sqrt{2 \zeta - 1}}{2}, \;  \zeta, \; \sqrt{\frac{1 \pm \sqrt{2 \zeta - 1}}{2}}
	\right),
	\label{eq:I:crossing:zeta}
\end{align}
of which the triangle vertex \eqref{eq:II:crossing:zeta} is the intersection point of edge II, and the triangle vertices \eqref{eq:I:crossing:zeta} are the two crossings of edge I with the constant $\zeta$ plane. Two-dimensional projections of the orbit space on the $\xi\zeta$-, $\xi\eta$- and $\zeta\eta$-planes are shown in Fig.~\ref{fig:xi:zeta}.

The minimum of the potential occurs on the convex hull of the orbit space \cite{Kim:1981xu,Heikinheimo:2017nth,Degee:2012sk}. Because the cross section of the orbit space is given by a triangle (the surface of the orbit space is a ruled surface), the convex hull is determined by the vertices and curved edges of the orbit space.

Only the neutral components of the Higgs doublet and triplet should obtain VEVs. Inserting these VEVs into the orbit space parameters, we find we must require that the global minimum be in the vertex $\vec{\rho}_{\text{A}} =  (1, 1, 1)$ of the orbit space. Electromagnetism is broken in the rest of the orbit space. For example, the charge-breaking extremum with $v_{h}/\sqrt{2} = v_{h^{+}}$, $v_{\delta^{++}} = -v_{\delta}$, and $v_{\delta^{+}} = 0$ considered in Ref.~\cite{Arhrib:2011uy} is given by $\vec{\rho} = (1/2, 1/2, 1/\sqrt{2})$ which lies on the end of edge IV where it meets edge I (but is not a vertex). Any extrema on other vertices and edges must have greater potential energy than that in vertex A. Moreover, because the edges III and IV are straight line segments, it is not necessary to consider them separately in the minimisation of the potential. They are automatically included in the convex hull of the orbit space by their end points.

%%%%%%%%%%%%%%%%%%%%%%%%%%%%%%%%%%%%%%
\section{Bounded-from-Below Conditions}
\label{sec:BfB}

It is known that if quartic terms in the scalar potential have a biquadratic $\lambda_{ij}\phi_i^2\phi_j^2$ form of real fields or gauge orbit variables, the potential is bounded from below if the $\lambda_{ij}$ matrix is copositive (positive on non-negative vectors) \cite{Cottle:1970,Kaplan:2000,Kannike:2016fmd}. However, for our potential in (\ref{eq:vpot}), a complication arises due to the last term which is not biquadratic. Note, though, that the constraints obtained neglecting the $\lambda_{SH\Delta}$ are necessary conditions for the potential to be BfB. 

In this work, we derive the BfB conditions in a scale-invariant singlet extention of type II seesaw model for the first time. They cannot be given in a fully analytical form, but can be found semi-numerically by solving the minimisation equations for the fields on a sphere, the Lagrange multiplier $\lambda$ enforcing that condition, and the orbit space variables. The details of the derivation and the necessary and sufficient conditions on the Higgs quartic couplings are given by in Appendix~\ref{app:BfB2}. 

%%%%%%%%%%%%%%%%%%%%%%%%%%%%%%%%%%%%%%

\section{Radiative Symmetry Breaking}\label{sec:RSB}

In multi-scalar theories, the treatment of radiative symmetry breaking requires a special care and general minimisation of the effective potential is a difficult task. A method of analysing the minimum of multi-scalar potential is devised by E. Gildener and S. Weinberg \cite{Gildener:1976ih}.  Since scalar couplings evolve with energy scale governed by their corresponding renormalisation group equations (RGEs),
the central idea of Gildener-Weinberg (GW) method is to choose a renormalisation scale $\mu_{\text{GW}}$ such that the tree-level potential develops a continuous line of degenerate non-trivial minima. Along this flat direction, even small loop corrections can change the shape of potential by developing a small curvature in the radial direction. In this sense, the GW method ensures a successful application of the Coleman-Weinberg radiative symmetry breaking mechanism in multi-scalar models.

%%%%%%%%%%%%%%%%%%%%%%%%%%%%%%%%%%%%%%
\subsection{Gildener-Weinberg Approach}

We now apply the Gildener-Weinberg method to our model. In the symmetry breaking vertex $\vec{\rho}_{\text{A}} =  (1, 1, 1)$ of the orbit space, the tree-level potential \eqref{eq:V:orbit} reads
\begin{equation}
\begin{split}
		V &= \lambda_{H} h^{4}+ (\lambda_{H\Delta}+\lambda_{H\Delta}^{\prime}) \delta^{2} h^{2}
		\\
		&+\lambda_{HS} s^{2} h^{2}+\lambda_{S} s^{4} + \lambda_{S\Delta} s^{2} \delta^{2}
		\\
		&+(\lambda_{\Delta}+\lambda_{\Delta}^{\prime}) \delta^{4}-\lambda_{SH\Delta}s\delta h^{2}.
\end{split}
\end{equation}
We now set all but the components that will get VEVs to zero, so the field norms are given by
\begin{equation}
  h^{2} = \frac{\phi_{h}^{2}}{2}, \quad s^{2} = \frac{S_{R}^{2}}{2}, \quad \delta^{2} = \frac{\phi_{\delta}^{2}}{2},
\end{equation}
and parametrise the fields as
\begin{equation}
		\phi_h=\varphi \, N_{h}, \quad S_{R}=\varphi \, N_{s}, \quad \phi_\delta=\varphi \, N_{\delta}.
\end{equation}
where $\varphi$ is the radial coordinate and $N_i$ has unit norm. At the scale $\mu_{\text{GW}}$, the tree-level potential admits a flat direction defined by $N_i = n_i$. The condition for the flat direction being a stationary line is given by considering the minimum with $V = 0$ on the unit sphere of fields, given by the stationary point equations
\begin{align}
		0&=\lambda_{H}n_{h}^{3}+\frac{1}{2}\left[(\lambda_{H\Delta}+\lambda_{H\Delta}^{\prime})n_{\delta}^{2}+\lambda_{HS}n_{s}^{2}\right]n_{h}
		\notag
		\\
		&-\frac{\lambda_{SH\Delta}}{2}n_{s}n_{\delta}n_{h},\label{eq:stat:h}\\\
		0&=(\lambda_{\Delta}+\lambda_{\Delta}^{\prime})n_{\delta}^{3}+\frac{1}{2}\left[ (\lambda_{H\Delta}+\lambda_{H\Delta}^{\prime})n_{h}^{2} \right. \notag \\ 
		&\left. + \lambda_{S\Delta}n_{s}^{2} \right]n_{\delta}-\frac{\lambda_{SH\Delta}}{4}n_{s}n_{h}^{2},\label{eq:stat:d}\\
		0&=\lambda_{S}n_{s}^{3}+\frac{1}{2}\left[\lambda_{S\Delta}n_{\delta}^{2}+\lambda_{HS}n_{h}^{2}\right]n_{s}
		\notag
		\\
		&-\frac{\lambda_{SH\Delta}}{4}n_{\delta}n_{h}^{2},\label{eq:stat:s}\\
		1&=n_{h}^{2}+n_{\delta}^{2}+n_{s}^{2}.
		\label{eq:unit}
\end{align}
		
Along the flat direction, a non-trivial minimum can be obtained by minimising the one-loop effective potential 
\begin{eqnarray} 		V_{\text{eff}}(\varphi)&=&A(\vec{n})\varphi^{4}+B(\vec{n})\varphi^{4}\log\frac{\varphi^{2}}{\mu_{\rm GW}^{2}}\label{1looppot}.
\end{eqnarray}
In the $\overline{\text{MS}}$ scheme, the dimensionless parameters $A(\vec{n})$ and $B(\vec{n})$ read
\begin{align}
		A(\vec{n})&=\frac{1}{64\pi^{2}v_{\varphi}^{4}}\left[6M_{W}^{4}\left(\log\frac{M_{W}^{2}}{v_{\varphi}^{2}}
 -\frac{5}{6}\right)
 \right.
 \notag
 \\
 & \left.
 +3M_{Z}^{4}\left(\log\frac{M_{Z}^{2}}{v_{\varphi}^{2}}-\frac{5}{6}\right)\right.\notag
 \\
		&\left.+\sum_{i}n_{i}M_{H_{i}}^{4}\left(\log\frac{M_{H_{i}}^{2}}{v_{\varphi}^{2}}-\frac{3}{2}\right)
		\right.
		\notag
		\\
		&\left. -12M_{t}^{4}\left(\log\frac{M_{t}^{2}}{v_{\varphi}^{2}}-\frac{3}{2}\right)\right],\\
		B(\vec{n})&=\frac{1}{64\pi^{2}v_{\varphi}^{4}}\Biggl[6M_{W}^{4}+3M_{Z}^{4}
		\notag
		\\
		& +\sum_{i}n_{i}M_{H_{i}}^{4} -12M_{t}^{4}\Biggr],
  \label{eq:B}
\end{align}
where the sum runs over the number of scalar mass eigenstates with $n_{i}=2$ for charged scalar and $n_{i}=1$ for neutral scalar. The  scalar Higgs mass spectrum after EWSB is provided in Section~\ref{sec:spectra}.

\subsection{Mass Spectrum}
\label{sec:spectra}

We calculate the scalar mass matrices and their mixing matrices. Note that in the end the mixing angles are completely determined by the flat direction components $n_{s}$, $n_{h}$ and $n_{\delta}$.

\subsubsection{Mass of the neutral CP-even Higgs}%c9o

The mass-squared matrix $\mathcal{M}^{2}_{R}$ of the neutral CP-even Higgs in the weak basis $(S_{R}, \phi_{h},\phi_{\delta})$ is given by
\begin{align}
  (\mathcal{M}^{2}_{R})_{11} &= \left[ 2\lambda_{S}n^{2}_{s}+\frac{\lambda_{SH\Delta}}{4}n^{2}_{h}\frac{n_{\delta}}{n_{s}} \right] v_{\varphi}^{2},
  \\
  (\mathcal{M}^{2}_{R})_{12} &= \left[\lambda_{HS}n_{h}n_{s}-\frac{\lambda_{SH\Delta}}{2}n_{h}n_{\delta} \right] v_{\varphi}^{2},
  \\
  (\mathcal{M}^{2}_{R})_{13} &= \left[\lambda_{S\Delta}n_{s}n_{\delta}-\frac{\lambda_{SH\Delta}}{4}n_{h}^2 \right] v_{\varphi}^{2},
  \\
  (\mathcal{M}^{2}_{R})_{22} &=  2\lambda_{H}n^{2}_{h}  v_{\varphi}^{2},
  \\
    (\mathcal{M}^{2}_{R})_{23} &=  \Bigl[ (\lambda_{H\Delta}+\lambda_{H\Delta}^{\prime})n_{h}n_{\delta}
    \notag
    \\
    &-\frac{\lambda_{SH\Delta}}{2}n_{h}n_{\delta} \Bigr] v_{\varphi}^{2},
    \\
   (\mathcal{M}^{2}_{R})_{33} &= \Bigl[ 2(\lambda_{\Delta}+\lambda_{\Delta}^{\prime})n^{2}_{\delta}
   \notag
   \\
   & +\frac{\lambda_{SH\Delta}}{4}n^{2}_{h}\frac{n_{s}}{n_{\delta}} \Bigr] v_{\varphi}^{2}.
\end{align}
The matrix $\mathcal{M}^{2}_{R}$ can be diagonalised by
\begin{align}
\mathcal{O}_{R}\;\mathcal{M}^{2}_{R}\;\mathcal{O}_{R}^{\text{T}}= \text{diag}\left(m_{\varphi}^{2} ,m_{h}^{2}, m_{H}^{2} \right).
\end{align}
The mixing matrix $\mathcal{O}_{R}$ is quite complicated, except for its first row that is given by the flat direction: 
\begin{equation}
  (\mathcal{O}_{R})_{1} = (n_{s}, n_{h}, n_{\delta}).
\end{equation}
The mass and weak eigenstates are related by 
\begin{equation}
\begin{pmatrix}
\varphi \\ h \\ H
\end{pmatrix}=\mathcal{O}_{R}
\begin{pmatrix}
S_{R} \\ \phi_{h} \\ \phi_{\delta}
\end{pmatrix}.
\end{equation}

\subsubsection{Mass of the neutral CP-odd Higgs}

The mass-squared matrix of the neutral CP-odd Higgs in the weak basis ($S_{I},\chi_{h},\chi_{\delta}$) is, after the minimum conditions are applied, given by
\begin{eqnarray}
\mathcal{M}^{2}_{I}=\frac{\lambda_{SH\Delta}}{2} v_{\varphi}^{2}
\begin{pmatrix}
\frac{n_{h}^{2}}{2}\frac{n_{\delta}}{n_{s}}& n_{h}n_{\delta}&-\frac{n_{h}^{2}}{2}\\\\ n_{h}n_{\delta}&2n_{s}n_{\delta}&-n_{h}n_{s}\\\\-\frac{n_{h}^{2}}{2}&-n_{h}n_{s}&\frac{n_{h}^{2}}{2}\frac{n_{s}}{n_{\delta}}
\end{pmatrix}.
\end{eqnarray}
The matrix rank of $\mathcal{M}^{2}_{I}$ is one and the null space of this matrix is two-dimensional. Hence, there are two massless fields: the unphysical Goldstone boson $G$ which will become the longitudinal component of the $Z$ boson, and the physical majoron $J$. 
The matrix $\mathcal{M}^{2}_{I}$ can be diagonalised by
\begin{align}
\mathcal{O}_{I} \, \mathcal{M}^{2}_{I}\;\mathcal{O}_{I}^{\text{T}}= \text{diag}\left(0,0,m_{A}^{2}\right),
\end{align}
where 
\begin{equation}
\mathcal{O}_{I} =
\begin{pmatrix}
   -C_{I1} n_{s} (n_{h}^{2} + 4 n_{\delta}^{2})
   &
   C_{I1} \, 2 n_{h} n_{\delta}^{2}
   &
   -C_{I1} n_{h}^{2} n_{\delta}
   \\
   0
   &
   C_{I2} \, n_{h}
   &
   C_{I2} \, 2 n_{\delta}
   \\
   -C_{I3} \, \frac{n_{\delta}}{n_{s}}
   &
   -C_{I3} \, \frac{2 n_{\delta}}{n_{h}}
   &
   C_{I3}
\end{pmatrix}
\label{eq:O:I}
\end{equation}
with
\begin{align}
  C_{I1}^{-1} &= \sqrt{(n_{h}^{2} + 4 n_{\delta}^{2})}
  \notag
  \\
  & \times \sqrt{(4 n_{s}^{2} n_{\delta}^{2} + n_{h}^{2} (n_{s}^{2} + n_{\delta}^{2}))},
  \\
  C_{I2}^{-1} &= \sqrt{n_{h}^{2} + 4 n_{\delta}^{2}},
  \\
  C_{I3}^{-1} &= \sqrt{1 + \left(\frac{4}{n_{h}^{2}} + \frac{1}{n_{s}^{2}} \right) n_{\delta}^{2}}.
\end{align}
%\begin{align}
%V&=\frac{1}{\sqrt{n_{h}^{2}+4n_{\delta}^{2}}},\notag\\
%a&=\frac{1}{\sqrt{{\frac{n_{s}^{2}}{V^{2}}+4n_{h}^{2}n_{\delta}^{4}}+n_{\delta}^{2}n_{h}^{4}}},\notag\\
%b&=\frac{1}{\sqrt{\frac{n_{h}^{2}}{V^{2}}+4n_{s}^{2}n_{\delta}^{2}}}.
%\end{align}

The mass and weak eigenstates are related by 
\begin{equation}
\begin{pmatrix}
J\\\ G\\A
\end{pmatrix}=\mathcal{O}_{I}\;\begin{pmatrix}
S_{I}\\\ \chi_{h}\\\chi_{\delta}
\end{pmatrix}.
\end{equation}

\subsubsection{Mass of the singly-charged Higgs}

The mass-squared matrix of the singly-charged Higgs is 
\begin{align}
\mathcal{M}^{2}_{\pm}&=
%\frac{\varphi^{2}}{2}\begin{pmatrix}
%\lambda_{SH\Delta}n_{s}n_{\delta}-\lambda_{H\Delta}^{\prime}n_{\delta}^{2}& \frac{\lambda_{H\Delta}^{\prime}}{\sqrt{2}}n_{h}n_{\delta}-\frac{\lambda_{SH\Delta}}{\sqrt{2}}n_{h}n_{s}\\\\ \frac{\lambda_{H\Delta}^{\prime}}{\sqrt{2}}n_{h}n_{\delta}-\frac{\lambda_{SH\Delta}}{\sqrt{2}}n_{h}n_{s}&\frac{\lambda_{SH\Delta}}{2}\frac{n_{s}}{n_{\delta}}n_{h}^{2}-\frac{\lambda_{H\Delta}^{\prime}}{2}n_{h}^{2}
%\end{pmatrix}\notag\\
%&=
\frac{v_{\varphi}^{2}}{4} (\lambda_{SH\Delta}n_{s} n_{h}-\lambda_{H\Delta}^{\prime} n_{\delta} n_{h})
\notag
\\
& \times \begin{pmatrix}
2\frac{n_{\delta}}{n_{h}}& -\sqrt{2}\\\\ -\sqrt{2}&\frac{n_{h}}{n_{\delta}}
\end{pmatrix}
\end{align}
in the weak basis $(h^{\pm},\delta^{\pm})$. The zero eigenvalue of $\mathcal{M}^{2}_{\pm}$ corresponds to the charged Goldstone boson absorbed by $W^{\pm}$. This mass matrix can be diagonalised by the orthogonal matrix $\mathcal{O}_{\pm}$ such that $\mathcal{O}_{\pm}\mathcal{M}^{2}_{\pm}\mathcal{O}_{\pm}^{\text{T}}=\text{diag}(m_{H^{\pm}},0)$, where
\begin{eqnarray}
\mathcal{O}_{\pm} 
=
%\begin{pmatrix}
%c_{\pm}& -s_{\pm}\\ s_{\pm}&c_{\pm}
%\end{pmatrix}=
\frac{1}{\sqrt{n_{h}^{2}+2n_{\delta}^{2}}}
\begin{pmatrix}
\sqrt{2}n_{\delta}&- n_{h}\\ n_{h}&\sqrt{2}n_{\delta}
\end{pmatrix},
\end{eqnarray}
and the physical charged Higgs mass is 
\begin{align}
m_{H^{\pm}}^{2}&=\frac{v_{\varphi}^{2}}{4}\left(\lambda_{SH\Delta}n_{s} n_{h}- \lambda_{H\Delta}^{\prime}n_{h}n_{\delta}\right)
 \frac{n_{h}^{2}+2n_{\delta}^{2}}{n_{h}n_{\delta}}.
\end{align}
\subsubsection{Mass of doubly charged Higgs}
%The mass squared of the doubly-charged Higgs is given by
%\begin{align}
%m_{H^{\pm\pm}}^{2}&= v_{\varphi}^{2}\left(\lambda_{\Delta}n_{\delta}^{2}+\frac{\lambda_{H\Delta}}{2}n_{h}^{2}+\frac{\lambda_{S\Delta}}{2}n_{s}^{2}\right).
%\end{align}
Applying the tadpole condition, the mass squared of the doubly-charged Higgs takes the form
\begin{align}
m^{2}_{H^{\pm\pm}}&=v_{\varphi}^{2}\left(\frac{\lambda_{SH\Delta}}{4}\frac{n_{s}}{n_{\delta}}n_{h}^{2}-\lambda_{\Delta}^{\prime}n_{\delta}^{2}-\frac{\lambda_{H\Delta}^{\prime}}{2}n_{h}^{2}\right).
\end{align}

%%%%%%%%%%%%%%%%%%%%%%%%%%%%%%%%%%%%%%
\subsection{Parametrisation via VEVs and Masses}

We now parametrise the quartic couplings via the VEVs of fields and the masses of particles. The scalar potential \eqref{eq:vpot} has nine free parameters; in addition, the flat direction component $n_{s}$ can be given via other ones. On the other hand, we have eight nonzero independent VEVs and masses: $v_{\varphi}$, $n_{h}$, $n_{\delta}$, $m_{A}$, $m_{h}$, $m_{H}$, $m_{H^{\pm}}$ and  $m_{H^{\pm\pm}}$.

We consider the tree-level mass hierarchy $m_{\varphi} < m_{h} < m_{H}$ of the CP-even mass eigenstates. We identify $h$ with the SM-like Higgs with $m_{h} = 125.25$~GeV. The mass of the dilaton $\varphi$ is zero at tree level (note that at one-loop level, the dilaton can become heavier than the SM-like Higgs).

We solve the Eqs. \eqref{eq:stat:h}, \eqref{eq:stat:d}, \eqref{eq:stat:s} and  \eqref{eq:unit} together with
\begin{align}
  m^{2}_{h} + m^{2}_{H} &= \tr \mathcal{M}^{2}_{R},
  \\
  m^{2}_{h} m^{2}_{H} &= \frac{1}{2} \Bigl[ (\tr \mathcal{M}^{2}_{R})^{2}
  \notag
  \\
  &  - \tr (\mathcal{M}^{2}_{R})^{2} \Bigr],
  \\ 
  m^{2}_{A} &= \tr \mathcal{M}^{2}_{I}, 
  \\
  m^{2}_{H^{\pm}} &= \tr \mathcal{M}^{2}_{\pm}, 
  \\
  m^{2}_{H^{\pm\pm}} &= \mathcal{M}^{2}_{\pm\pm}, 
\end{align}
where we take into account that the dilaton mass is zero at tree level and that $\mathcal{M}^{2}_{I}$ and $\mathcal{M}^{2}_{\pm}$ also contain zero eigenvalues -- Goldstone masses.\footnote{We use the three invariants  of a $3 \times 3$ matrix $\mathcal{M}^{2}$ in terms of its eigenvalues $m^{2}_{i}$, i.e. $\tr \mathcal{M}^{2} = m_{1}^{2} + m_{2}^{2} + m_{3}^{2}$, $\tr (\operatorname{adj} \mathcal{M}^{2}) = \frac{1}{2} [ (\tr \mathcal{M}^{2})^{2} - \tr (\mathcal{M}^{2})^{2} ] = m^{2}_{1} m^{2}_{2} + m^{2}_{1} m^{2}_{3} + m^{2}_{2} m^{2}_{3}$ and  $\det \mathcal{M}^{2} = m_{1}^{2} m_{2}^{2} m_{3}^{2}$.} Note that the equations $\det \mathcal{M}^{2}_{R} = \det \mathcal{M}^{2}_{I} = \det \mathcal{M}^{2}_{\pm} = 0$ do not provide further constraints on quartic couplings.

Considering, without loss of generality, only $n_{s} > 0$, the system of equations has two solutions, of which we pick the one that tends to give perturbative values to quartic couplings. Because we have nine free parameters in the potential, but eight VEVs and masses, we have to specify the value of one of the quartic couplings. For this we choose $\lambda_{\Delta}$, because it is more convenient to remain within perturbativity bounds in this way. Unfortunately the solutions to the above equations are too lengthy to present explicitly. The solutions for $\lambda'_{\Delta}$ and $\lambda_{H\Delta}$ are sensitive to the value of the triplet VEV and their expansion in Taylor series results in inaccurate expressions.

The dilaton mass $m_{\varphi}$ arises at one-loop level via
\begin{equation}
  m_{\varphi} = 8 B(\vec{n})
\end{equation}
with $B(\vec{n})$ given by Eq.~\eqref{eq:B}. All the mixing angles of the mass matrices are also determined by the VEVs and masses. In particular, since $\varphi$ is the scalon, the first row of the CP-even scalar mixing matrix is given by the flat direction unit vector $\vec{n}$.

%%%%%%%%%%%%%%%%%%%%%%%%%%%%%%%%%%%%%%%
%\subsection{Coleman-Weinberg Approximation}
%
%In the simplest approximation, only the $S$ field obtains a VEV via the Coleman-Weinberg mechanism. Subsequently, the VEV of $S$ induces dimensionful couplings for the Higgs and the triplet, which generate VEVs for them.
%The correspondence between the scale-invariant potential parameters and the parameters of the type II seesaw are given by
%\begin{align}
%  \mu^{2}_{H} &= \lambda_{HS} v_{s}^{2},
%  \label{eq:CW:mu:H:sq}
%  \\
%  \mu^{2}_{\Delta} &= \lambda_{S\Delta} v_{s}^{2},
%  \label{eq:CW:mu:D:sq}
%  \\
%  \mu &= \lambda_{SH\Delta} v_{s}.
%  \label{eq:CW:mu}
%\end{align}
%Note that this can only be an approximation, since a negative $\lambda_{HS}$ yields a negative mass matrix eigenvalue if only $S$ acquires a VEV: the true minimum is off $S$ axis. In particular, this approximation neglects mixing of other fields with $S$. Given the correspondence in Eqs. \eqref{eq:CW:mu:H:sq}, \eqref{eq:CW:mu:D:sq} and \eqref{eq:CW:mu}, the formulae of \cite{Arhrib:2011uy}, relating the couplings of the type II seesaw potential to physical parameters, all apply. In addition to the field content of the type II seesaw, there are two new particles, the scalon and the majoron. The scalon mass, using the flat direction condition $\lambda_{S} = 0$, is given by
%\begin{equation}
%  m_{s}^{2} = \frac{2 \lambda_{HS}^{2} + 3 \lambda_{S\Delta}^{2}}{16 \pi^{2}} v_{s}^{2},
%\end{equation}
%while the majoron is massless.
%
%\KK{ In which parts of the parameter space is this approximation bad?
%}

%%%%%%%%%%%%%%%%%%%%%%%%%%%%%%%%%%%%%
\section{Numerical Study}
\label{sec:result}

%%%%%%%%%%%%%%%%%%%%%%%%%%%%%%%%%%%%%%%
\begin{figure*}[tb]
	\begin{center}
		\includegraphics{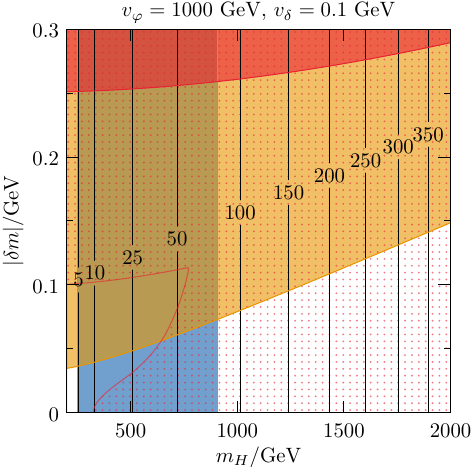}~\includegraphics{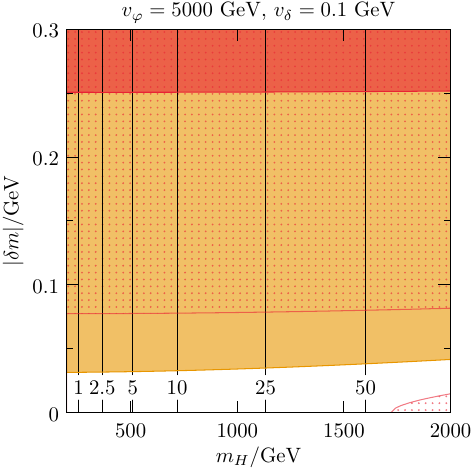}
		\caption{The parameter space on the $\abs{\delta m}$  vs. $m_{H}$ plane with $v_{\delta} = 0.1$~GeV and $\lambda_{\Delta} = 0.1$. In the left panel, $v_{\varphi} = 1000$~GeV; in the right panel, $v_{\varphi} = 5000$~GeV. The black lines are contours of the dilaton mass $m_{\varphi}/\text{GeV}$. The couplings are non-perturbative in the red region (not perturbative up to the Planck scale in the red dotted region) and the potential is not bounded from below in the yellow region. The blue region is forbidden by the mixing of the Higgs boson with the other fields.}
		\label{fig:deltam:mH}
	\end{center}
\end{figure*}
%%%%%%%%%%%%%%%%%%%%%%%%%%%%%%%%%%%%%%%

In this section, we show representative examples of the parameter space with radiative symmetry breaking that results in the electroweak vacuum together with various theoretical and experimental constraints.

We fix the doublet and triplet Higgs VEVs to the combination $v\equiv\sqrt{v_{h}^{2}+2v_{\delta}^{2}}=246.22$~GeV and the SM-like Higgs mass $m_h=125.25$~GeV. For the triplet self-coupling $\lambda_{\Delta}$, we use the value $\lambda_{\Delta} = 0.1$ which is sufficient to ensure vacuum stability but not too large so as not to run non-perturbative at a high scale.

We consider the following experimental and theoretical constraints on the parameter space:
\begin{itemize}
  \item The $\rho$ parameter;
  \item Electroweak precision parameters;
  \item Collider bounds on $H^{++}$;
  \item Energy loss from red giants via Majorons;
  \item Mixing of the Higgs boson with other scalars;
  \item Higgs-to-Majoron decay $h \to J J$;
  \item Higgs-to-dilaton decay $h \to \varphi \varphi$;
  \item Bounded-from-below conditions;
  \item Perturbativity of the quartic couplings in the minimum;
  \item Perturbativity of couplings at the Planck scale.
\end{itemize}
As usual with a mostly singlet majoron, the $Z \to \varphi J$ decay is negligible and does not constrain the parameter space.

Because in the type II seesaw model the triplet component masses commonly lie near a common mass scale, we define as usual
\begin{equation}
  \delta m_{1} = m_{H^{\pm}} - m_{H}, \quad \delta m_{2} = m_{H^{\pm \pm}} - m_{H^{\pm}}.
\end{equation}
The triplet VEV contributes to the $\rho$ parameter $\rho \equiv m_{W}^{2}/(m_{Z}^{2} c_{W}^{2})$, where $c_{W}$ is the cosine of the Weinberg angle. Comparing the value $\rho = 1.00038 \pm 0.00020$ from a global fit \cite{ParticleDataGroup:2020ssz} with $\rho \approx 1 - v_{\delta}^{2}/v^{2}$ from the type II seesaw, one obtains the bound $v_{\delta} \leq 2.6$~GeV at the $3 \sigma$~C.L. \cite{Mandal:2022zmy}. The mass differences of the triplet components cannot be arbitrarily large due to constraints from the electroweak precision parameters \cite{Peskin:1991sw,Peskin:1990zt}. From a global fit on the $S$ and $T$ parameter (with $U = 0$) \cite{ParticleDataGroup:2020ssz}, one obtains $\abs{\delta m_{1}} \approx \abs{\delta m_{2}} \leq 45.5$~GeV at $90\%$~C.L. \cite{Mandal:2022zmy}. In our examples, we take $\delta m_{1} = \delta m_{2} = \delta m$ and $m_{A} = m_{H}$.

The doubly-charged scalar decays predominantly into gauge bosons for $v_{\delta} > 10^{-4}$~GeV, giving the bound on its mass $m_{H^{++}} \geq 220$~GeV, while for $v_{\delta} < 10^{-4}$~GeV, one has $m_{H^{++}} \geq 870$~GeV since then it will decay predominantly into leptons \cite{Melfo:2011nx}.

A strong constraint on the pseudoscalar mixing comes from the energy loss from red giant stars via the process $\gamma + e^{-} \to e^{-} + J$, since the Majoron can escape the star \cite{Georgi:1981pg,Choi:1989hi,Montero:2011jk,Sanchez-Vega:2014rka}. This restricts the coupling
\begin{equation}
  g_{\bar{e}eJ} = \frac{y_{e}}{\sqrt{2}} (\mathcal{O}_{I})_{12} \approx \frac{2 m_{e}}{v_{h}^{2}} \frac{v_{\delta}^{2}}{v_{s}}
\end{equation}
to be within $g_{\bar{e}eJ} \leq 10^{-10}\text{ to }10^{-12}$. Since the $g_{\bar{e}eJ}$ coupling is suppressed by $v_{\delta}^{2}$, this constraint only requires $v_{\delta} \leq 10^{-1}$~GeV in order to be satisfied.

The mixing of the Higgs boson with other CP-even fields, given by the $\abs{(\mathcal{O}_{R})_{22}}$ element of the CP-even mixing matrix, is constrained by global fits of the Higgs couplings and by the LEP data \cite{Robens:2016xkb}.

If the dilaton mass is less than $m_{h}/2$, then the SM-like Higgs boson can decay into dilatons
 with the decay width
\begin{equation}
  \Gamma_{h \to \varphi \varphi} = \frac{g_{h\varphi\varphi}^{2}}{32 \pi m_{h}} 
  \sqrt{1 - \frac{4 m_{\varphi}^{2}}{m_{h}^{2}}}.
\end{equation}
If the branching ratio $\text{BR}_{h \to \varphi \varphi}$ is large enough, this significantly constrains the Higgs mixing \cite{Robens:2015gla}.

The $h \to J J$ decay will contribute to the Higgs invisible width. The decay width is given by
\begin{equation}
  \Gamma_{h \to JJ} = \frac{1}{32 \pi} \frac{g_{hJJ}^{2}}{m_{h}},
\end{equation}
while the SM Higgs width is $\Gamma_{h \to \text{SM}} = 4.07 \times 10^{-3}$~GeV. The Higgs invisible branching ratio is given by
\begin{equation}
  \text{BR}_{h \to \text{inv}} = \frac{\Gamma_{h \to JJ} + \Gamma_{h \to \varphi\varphi} \text{BR}^2_{\varphi \to JJ}}{\Gamma_{h \to \text{SM}} + \Gamma_{h \to \varphi\varphi} + \Gamma_{h \to JJ}},
  \label{eq:h:to:JJ}
\end{equation}
where 
\begin{equation}
  \text{BR}_{\varphi \to JJ} = \frac{\Gamma_{\varphi \to JJ}}{\Gamma_{h \to \text{SM}}(m_{\varphi}) (\mathcal{O}_{R})_{12}^{2} + \Gamma_{\varphi \to JJ}}.
\end{equation}
We have $(\mathcal{O}_{R})_{12} = n_{h}$ and $\Gamma_{h \to \text{SM}}(m_{\varphi})$ is obtained from \cite{Gomez-Bock:2007azi}.\footnote{Numerically, the second term in the numerator of Eq. \eqref{eq:h:to:JJ} is negligible. We also neglect the contribution of the triplet component of the dilaton to the decay with into the SM, since it is proportional to $n_{\delta}^{2}$.}
Latest measurements by the CMS experiment at the LHC find $\text{BR}_{h \to \text{inv}}< 0.18$ \cite{CMS:2022qva}, while the ATLAS experiment finds $\text{BR}_{h \to \text{inv}}< 0.145$ \cite{ATLAS:2022yvh}; we require the latter constraint.

We have also identified the parameter space in which the couplings remain perturbative up to the Planck scale, by calculating the RG running with the RGEs given in Appendix~\ref{app:rges}. As initial values of gauge couplings and top Yukawa coupling, we use $g_{Y}(M_{t})=0.35745$, $g_{2}(M_{t}) =0 .64779$, $g_{3}(M_{t}) = 1.1666$, $y_{t}(M_{t}) = 0.93690$~\cite{Buttazzo:2013uya}.

We also comment on the fate of the Higgs doublet quartic coupling from weak scale to Planck scale. As is well known that RGE running of quartic coupling in SM crosses zero around $10^{10}$~GeV due to the strong negative contribution from the top Yukawa term \cite{Degrassi:2012ry, Buttazzo:2013uya}. The situation can be dramatically changed with positive contributions from additional bosons. In the case of singlet extension of type II seesaw, there are new contributions to the the Higgs quartic $\beta$-function from the portal couplings $\lambda_{H\Delta}$, $\lambda_{H\Delta}^{\prime}$, and $\lambda_{HS}$. It can be seen that in this model, the $\lambda_{H}$ can remain positive up to the Planck scale signaling that the vacuum will be stable.

The parameter space in the $\abs{\delta m}$  vs. $m_{H}$ plane is shown in Figure~\ref{fig:deltam:mH}.\footnote{The parameter space is practically symmetric in $\delta m$ for the range of parameters we show, so we only show positive $\abs{\delta m}$; in larger regions this may not hold.} The couplings are non-perturbative at the weak scale in the red region and not perturbative up to the Planck scale in the dotted red region (in this region, a Landau pole arises at the scale $10^{8}$~GeV at the highest). The potential is not bounded from below in the yellow region. Both the BfB and non-perturbativity bounds arise from $\lambda'_{\Delta}$ that becomes large and negative with larger $\abs{\delta m}$. The BfB bound is mostly due to violation of the $\lambda_{\Delta} + \lambda'_{\Delta} > 0$ condition in  Eq.~\eqref{eq:BfB:BC}. The blue region is forbidden by the mixing of Higgs and other scalars which alters the Higgs signal strengths \cite{Robens:2016xkb}. The left panel of Figure~\ref{fig:deltam:mH} shows the parameter space for $v_{\varphi} = 1000$~GeV; in the right panel, $v_{\varphi} = 5000$~GeV, while $v_{\delta} = 0.1$~GeV in both cases. For $v_{\varphi} = 1000$~GeV, only the lower-left corner of the plot presents parameter space that satisfies all the constraints. For the larger $v_{\varphi} = 5000$~GeV, the CP-even scalar mixing is not constraining and the couplings remain perturbative up to the Planck scale in a larger region.

%%%%%%%%%%%%%%%%%%%%%%%%%%%%%%%%%%%%%%%
\begin{figure}[t]
	\begin{center}
		\includegraphics{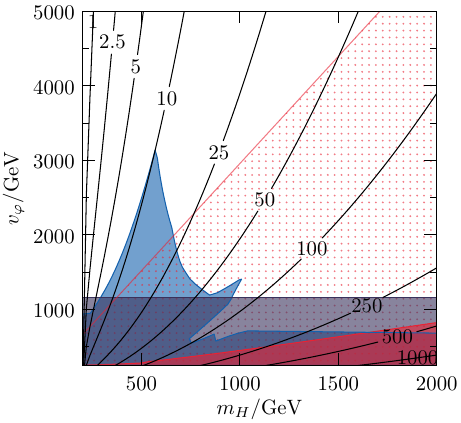}
		\caption{The parameter space on the $v_{\varphi}$  vs. $m_{H}$ plane with $\delta m = 0$~GeV. The black lines are contours of the dilaton mass $m_{\varphi}/\text{GeV}$. The couplings are non-perturbative in the red region (not perturbative up to the Planck scale in the red dotted region). The blue region is forbidden by the mixing of other scalars with the Higgs boson and the violet region by the Higgs invisible width from Higgs decay into majorons.}
		\label{fig:vvarphi:mH}
	\end{center}
\end{figure}
%%%%%%%%%%%%%%%%%%%%%%%%%%%%%%%%%%%%%%%

Because in most cases, as we see, the mass difference $\delta m$ has to be very small, it is interesting to study separately the parameter space with $\delta m = 0$. This is shown in Fig.~\ref{fig:vvarphi:mH} in the $v_{\varphi}$  vs. $m_{H}$ plane with contours of the dilaton mass $m_{\varphi}$ (black lines). This plot is valid for any small value of $v_{\delta}$. The quartic couplings are non-perturbative in the solid red region and have a Landau pole $\Lambda < m_{P}$ in the dotted red region. The blue region is forbidden by the mixing of other scalars with the Higgs boson which alters the Higgs signal strengths and the violet region by the Higgs invisible branching ratio Eq.~\eqref{eq:h:to:JJ}. The branching ratio $\text{BR}_{h \to \varphi \varphi}$ practically vanishes for the shown parameter space. Satisfying other constraints (except perturbativity up to the Planck scale), with $v_{\varphi} = 600$~GeV, the Higgs quartic can be down to $83\%$ of its SM value. When perturbativity up to the Planck scale is required, the value differs from the SM value up to $5\%$. The Higgs quartic remains positive up to the Planck scale in the same region in which couplings remain perturbative up to the Planck scale.

%%%%%%%%%%%%%%%%%%%%%%%%%%%%%%%%%%%%%%%
\begin{table*}[t]
	\begin{center}
		\caption{A few benchmark points with $\delta m = 0$~GeV, $v_{\delta} = 0.1$~GeV and $\lambda_{\delta} = 0.1$ that satisfy all constraints.} 
			\label{tab:quarticcoup}
	\footnotesize
	\begin{tabular}{lllllllllllll}
		\toprule BP&
		$m_{H}/\text{GeV}$ &
		$v_{\varphi}/\text{GeV}$ &
		$v_\delta$ &
		$\lambda_{H}$ &
		$\lambda_{\Delta}^{\prime}$&
		$\lambda_{S}$&
		$\lambda_{HS}$&
		$\lambda_{H\Delta}$&
		$\lambda_{H\Delta}^{\prime}$&
		$\lambda_{S\Delta}$&
		$\lambda_{SH\Delta}$\\
		\midrule
		A & $225$ & $1500$ & $0.1$ & $0.126$ & $0.023$ & $9.7 \times 10^{-4}$ & $-0.00697$ & $0.330$ & $-1.11 \times 10^{-6}$ & $0.037$ & $2.3 \times 10^{-4}$ \\
		B & $225$ & $5000$ & $0.1$ & $0.129$ & $0.0020$ & $7.6 \times 10^{-7}$ & $-6.28 \times 10^{-4}$ & $0.325$ & $-1.10 \times 10^{-8}$ & $0.0033$ & $6.6 \times 10^{-5}$ \\
		C & $1000$ & $5000$ & $0.1$ & $0.129$ & $0.040$ & $7.6 \times 10^{-7}$ & $-6.28 \times 10^{-4}$ & $0.329$ & $-2.18 \times 10^{-7}$ & $0.0079$ & $1.3 \times 10^{-3}$ \\
		\botrule
	\end{tabular}
\end{center}
\end{table*}
%%%%%%%%%%%%%%%%%%%%%%%%%%%%%%%%%%%%%%%

As an example, the values of quartic couplings for three benchmark points that satisfy all constraints are listed in Table~\ref{tab:quarticcoup}. Point A is chosen with a small $m_{H} = 225$~GeV in the region where a $v_{\varphi} = 1.5$~TeV is allowed: in this point, $\lambda_{H} = 0.126$ is smaller than its SM value. In points B and C, we choose a larger $v_{\varphi} = 5$~TeV and the Higgs quartic coupling is practically the same as in the SM.

\begin{figure*}[tb]
	\begin{center}
		\includegraphics{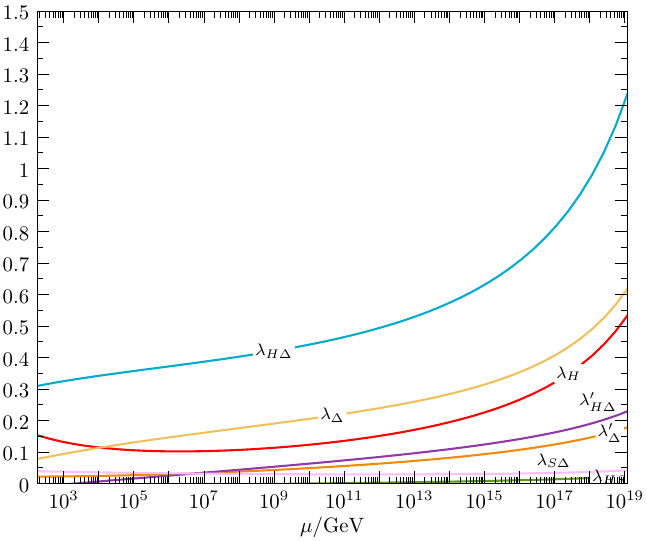}
		\caption{RGE running of scalar quartic couplings for the benchmark point A in Table~\ref{tab:quarticcoup}. The tiny couplings $\lambda_{S}$ and $\lambda_{HS\Delta}$ that run very little are not shown.}
		\label{fig:rge:run:A}
	\end{center}
\end{figure*}

RGE running of scalar quartic couplings for the benchmark point A in Table~\ref{tab:quarticcoup} is demonstrated in Figure~\ref{fig:rge:run:A}. The couplings $\lambda_{S}$ and $\lambda_{HS\Delta}$ that are tiny and run very little are not shown. The RGE running for the other two benchmark points is rather similar.

%%%%%%%%%%%%%%%%%%%%%%%%%%%%%%%%%%%%%%%
%%%%%%%%%%%%%%%%%%%%%%%%%%%%%%%%%%%%%%%

%%%%%%%%%%%%%%%%%%%%%%%%%%%%%%%%%%%%%
\newpage
\section{Conclusions}
\label{sec:conclusions}

In this paper, we have considered the singlet extension of type II seesaw possessing classical scale invariance. A new scalar singlet has been introduced, whose VEV spontaneously breaks the global lepton number symmetry. Consequently, the majoron -- the Goldstone boson of lepton number breaking -- is mostly singlet-like. This framework is interesting in three aspects. First, the triplet Yukawa coupling of type II seesaw, together with spontaneous breaking of the lepton number, addresses the neutrino mass problem. Second, a classical scale-invariant theory paves the way to the origin of the electroweak potential which also allows us to cure the hierarchy problem. Last, the incorporation of a new bosonic degree of freedom can save the vacuum of the theory from being unstable.
		
In order to minimise a complicated scalar potential, we determine and use the gauge orbit space of the model. A full set of sufficient and necessary conditions for the scalar potential to be bounded from below is derived in Appendix \ref{app:BfB2}. The multi-scalar potential is minimised with the Gildener-Weinberg method. The quartic couplings are parametrised in terms of VEVs and masses.
% An important role of the radiative corrections of the tree-level potential can be at play along the flat direction. At the scale $\varphi=1$~TeV, the effective potential develops a non-zero VEV signaling that the symmetry has been spontaneously broken. In this regard, the scale has been generated dynamically by quantum corrections unlike the prior given scale in the spontaneously breaking of electroweak symmetry in the SM.
		
We showed that the perturbativity of quartic couplings and the stability of electroweak vacuum can be maintained all the way up to the Planck scale with the new contributions coming from the singlet and triplet scalars. In particular, the evolution of $\lambda_{H}$ with the energy scale can be prevented from crossing zero value at high energy due to sizeable contributions from $\lambda_{H\Delta}$ and $\lambda_{H\Delta}^{\prime}$. In the allowed parameter space, demonstrated in Figures~\ref{fig:deltam:mH} and \ref{fig:vvarphi:mH}, the mass splittings between triplet-like states have to be almost zero.
		
In conclusion, we have shown in this work that radiative symmetry breaking can be realised in the scale-invariant singlet extension of type II seesaw model, taking into account restrictions from collider experiments and astrophysics. Due to new scalar fields, the model has rich phenomenology. 

\backmatter

%\bmhead{Supplementary information}

\bmhead{Acknowledgments}

BD and WS acknowledge support from Suranaree University of Technology (SUT). BD was supported by Thailand Science Research and Innovation and Suranaree University of Technology through SUT-Ph.D. Scholarship Program for ASEAN. KK was supported by the Estonian Research Council grant PRG434, by the European Regional Development Fund and the programme Mobilitas Pluss grant MOBTT5, and by the EU through the European Regional Development Fund CoE program TK133 ``The Dark Side of the Universe''. 

%\section*{Declarations}
%
%Some journals require declarations to be submitted in a standardised format. Please check the Instructions for Authors of the journal to which you are submitting to see if you need to complete this section. If yes, your manuscript must contain the following sections under the heading `Declarations':
%
%\begin{itemize}
%\item Conflict of interest/Competing interests (check journal-specific guidelines for which heading to use)
%\item Ethics approval 
%\item Consent for publication
%\item Availability of data and materials
%\item Authors' contributions
%\end{itemize}
%
%\noindent
%If any of the sections are not relevant to your manuscript, please include the heading and write `Not applicable' for that section. 
%
%%%===================================================%%
%%% For presentation purpose, we have included        %%
%%% \bigskip command. please ignore this.             %%
%%%===================================================%%
%\bigskip
%\bigskip\noindent
%Springer journals and proceedings: \url{https://www.springer.com/gp/editorial-policies}
%
%\bigskip\noindent
%Nature Portfolio journals: \url{https://www.nature.com/nature-research/editorial-policies}
%
%\bigskip\noindent
%\textit{Scientific Reports}: \url{https://www.nature.com/srep/journal-policies/editorial-policies}
%
%\bigskip\noindent
%BMC journals: \url{https://www.biomedcentral.com/getpublished/editorial-policies}
%\end{flushleft}
%
\begin{appendices}

%%%%%%%%%%%%%%%%%%%%%%%%%%%%%%%%%%%%%%%
%%%%%%%%%%%%%%%%%%%%%%%%%%%%%%%%%%%%%%%
\section{Derivation of Bounded-from-Below Conditions}
\label{app:BfB2}

We derive the necessary and sufficient bounded-from-below conditions for the scalar potential. Because the potential \eqref{eq:V:orbit} is linear in the orbit space variables, its minimum with respect to them lies on the boundary of the orbit space, more precisely on the intersection of the boundary and its convex hull. As discussed at the end of Sec.~\eqref{sec:orbit:space}, it is enough to give the conditions at the vertices A, B and C \eqref{eq:vertices} and at the edges I \eqref{eq:edge:I} and II \eqref{eq:edge:II} of the orbit space. Notice that the end points of edge I are vertices A and C, and the end points of edge II are vertices B and C. If there are no physical solution inside an edge, then the edge minimum is at an end point.

The vertex A is already accounted for, because we require the flat direction of the potential to lie there. At vertices B and C and edge II, the orbit space parameter $\eta = 0$ which makes the potential biquadratic there. Therefore at B and C we can derive BfB conditions by requiring copositivity of the quartic coupling matrix \cite{Kannike:2012pe}:
\begin{equation}
  \Lambda = 
  \begin{pmatrix}
  \lambda_{H} & \frac{1}{2} (\lambda_{H\Delta} + \xi \lambda'_{H\Delta}) & \frac{1}{2} \lambda_{HS}
  \\
  \frac{1}{2} (\lambda_{H\Delta} + \xi \lambda'_{H\Delta}) & \lambda_{\Delta} + \zeta \lambda'_{\Delta} 
  & \frac{1}{2} \lambda_{S\Delta}
  \\
  \frac{1}{2} \lambda_{HS} & \frac{1}{2} \lambda_{S\Delta} & \lambda_{S}
  \end{pmatrix}.
  \label{eq:Lambda:matrix}
\end{equation}
The copositivity conditions for the matrix \eqref{eq:Lambda:matrix} read
\begin{equation}
\begin{split}
  \lambda_{H} &> 0, 
  \quad 
  \lambda_{\Delta} + \zeta \lambda'_{\Delta} > 0, 
  \quad
  \lambda_{S} >0,
  \\
  \bar{\lambda}_{H\Delta} &\equiv \frac{1}{2} (\lambda_{H\Delta} + \xi \lambda'_{H\Delta})
  \\
  & + \sqrt{\lambda_{H} (\lambda_{\Delta} + \zeta \lambda'_{\Delta})} > 0,
  \\
  \bar{\lambda}_{HS} \equiv& \frac{1}{2} \lambda_{HS} + \sqrt{\lambda_{H} \lambda_{S}} > 0,
  \\
   \bar{\lambda}_{S\Delta} &\equiv \frac{1}{2} \lambda_{S\Delta} + \sqrt{\lambda_{S} (\lambda_{\Delta} + \zeta \lambda'_{\Delta})},
   \\
  & \sqrt{\lambda_{H} (\lambda_{\Delta} + \zeta \lambda'_{\Delta}) \lambda_{S}}  
   + \frac{1}{2} \lambda_{S\Delta} \sqrt{\lambda_{H}}
  \\
  & + \frac{1}{2} \lambda_{HS} \sqrt{\lambda_{\Delta} + \zeta \lambda'_{\Delta}}
  \\
  &+ \frac{1}{2} (\lambda_{H\Delta} + \xi \lambda'_{H\Delta}) \sqrt{\lambda_{S}}
  \\
  & + \sqrt{2 \bar{\lambda}_{H\Delta} \bar{\lambda}_{HS} \bar{\lambda}_{S\Delta}} > 0.
\end{split}
\label{eq:BfB:BC}
\end{equation}
These conditions must hold true for the values of orbit space variables $\xi$ and $\zeta$ at both vertices B and C \eqref{eq:vertices}.

On edge II, we can minimise the potential \eqref{eq:V:orbit} on a unit sphere of fields together with the orbit variable $\xi$ parametrising the edge and the Lagrange multiplier $\lambda$ by solving
\begin{equation}
\begin{split}
  2 \lambda s &= s \, (2 \lambda_{HS} h^{2} + 4 \lambda_{S} s^{2} + 2 \lambda_{S\Delta} \delta^{2}),
  \\
  2 \lambda h &= h \, [2 \lambda_{HS} s^{2} + 4 \lambda_{H} h^{2} 
  \\
  &+ 2 (\lambda_{H\Delta} + \xi \lambda'_{H\Delta}) \delta^{2}],
  \\
  2 \lambda \delta &= \delta [2 \lambda_{S\Delta} s^{2} + 2 (\lambda_{H\Delta} 
  + \xi \lambda'_{H\Delta}) h^{2}
  \\
  &+ 4 (\lambda_{\Delta} + (1- 2 \xi^{2}) \lambda'_{\Delta}) \delta^{2}],
  \\
  0 &= \lambda'_{H\Delta} h^{2} \delta^{2} - 4 \xi \lambda'_{\Delta} \delta^{4},
  \\
  1 &= h^{2} + s^{2} + \delta^{2}.
\end{split}
\label{eq:edge:II:BfB:eqs}
\end{equation}
These equations can be solved analytically. For each solution, one has to check whether the variables are in the physically allowed range and if they are, check that the Lagrange parameter $\lambda$,  proportional to the potential $V$ for this solution, is greater than zero:
\begin{equation}
\begin{split}
  0 &< h^{2} < 1 \land 0 \leq s^{2} < 1 \land 0 < \delta^{2} < 1  \\
  & \land 0 < \xi < \frac{1}{2} \implies V > 0.
\end{split}
\label{eq:edge:II:BfB:conds}
\end{equation}
Notice that $p \implies q$ is equivalent to $\lnot p \lor q$ and also that $\lambda \propto V$ for each solution.

On edge I, the minimisation equations for the fields on a unit sphere, $\xi$ and $\lambda$ are given by
\begin{equation}
\begin{split}
  2 \lambda s &= s \, (2 \lambda_{HS} h^{2} + 4 \lambda_{S} s^{2} + 2 \lambda_{S\Delta} \delta^{2})
  \\
  &- \sqrt{\xi}\abs{ \lambda_{SH\Delta}} h^{2} \delta,
  \\
  2 \lambda h &= h \, [2 \lambda_{HS} s^{2} + 4 \lambda_{H} h^{2} 
  \\
  &+ 2 (\lambda_{H\Delta} + \xi \lambda'_{H\Delta}) \delta^{2} - 2 \sqrt{\xi}\abs{ \lambda_{SH\Delta}} s \delta],
  \\
  2 \lambda \delta &= \delta [2 \lambda_{S\Delta} s^{2} + 2 (\lambda_{H\Delta} 
  + \xi \lambda'_{H\Delta}) h^{2}
  \\
  &+ 4 (\lambda_{\Delta} + (1 - 2 \xi + 2 \xi^{2}) \lambda'_{\Delta}) \delta^{2}]
  \\
  &- \sqrt{\xi}\abs{ \lambda_{SH\Delta}} h^{2} s,
  \\
  0 &= h^{2} \left(\lambda'_{H\Delta} \delta^{2} - \frac{\abs{ \lambda_{SH\Delta}} s \delta}{2 \sqrt{\xi}} \right)
  \\
  &+ 2 (2 \xi - 1) \lambda'_{\Delta} \delta^{4},
  \\
  1 &= h^{2} + s^{2} + \delta^{2}.
\end{split}
\label{eq:edge:I:BfB:eqs}
\end{equation}
These equations can only be solved numerically.\footnote{Because usually $\lambda_{SH\Delta}$ is very small, good necessary conditions are obtained by setting it to zero in Eq. \eqref{eq:edge:I:BfB:eqs}.} Similarly to the case of Eq. \eqref{eq:edge:II:BfB:conds} for edge II, one has to check that the solutions are in the physical range before checking that $V > 0$ with these arguments:
\begin{equation}
\begin{split}
  0 &\leq h < 1 \land 0 \leq s < 1 \land 0 \leq \delta < 1  \\
  & \land 0 < \xi < 1 \implies V > 0.
\end{split}
\label{eq:edge:I:BfB:conds}
\end{equation}
Altogether, since vertex A is accounted for by the requirement of a flat direction, the BfB conditions can be written as
\begin{equation}
  V\rvert_{\rm B} > 0 \land V\rvert_{\rm C} > 0 \land V\rvert_{\rm II} > 0 \land V\rvert_{\rm I} > 0,
\end{equation}
where the first two conditions are given by Eq. \eqref{eq:BfB:BC} with, respectively, the values of the orbit variables at vertices B and C inserted, and the last two conditions are given by Eq. \eqref{eq:edge:II:BfB:conds} which has to be satisfied for each solution of Eq. \eqref{eq:edge:II:BfB:eqs} and \eqref{eq:edge:I:BfB:conds} which has to be satisfied for each solution of Eq. \eqref{eq:edge:I:BfB:eqs}.

%%%%%%%%%%%%%%%%%%%%%%%%%%%%%%%%%%%%%%%
%%%%%%%%%%%%%%%%%%%%%%%%%%%%%%%%%%%%%%%
\section{RGEs of quartic couplings}
\label{app:rges}

We use the PyR@TE package \cite{Sartore:2020gou} to calculate the beta-functions of all scalar quartic couplings, gauge couplings and the top Yukawa coupling at two-loop level (we have ignored all other Yukawa couplings). For conciseness, we only provide the one-loop results here, while in our numerical study we use the two-loop beta-functions. The beta-functions are given by
\begin{align}
	%%%%%%%%%%%%%%\lambda_{H}
		\frac{\text{d}\lambda_{H}}{\text{d}t}&=\frac{1}{16\pi^2}\Bigr[24\lambda_{H}^{2}+\frac{1}{2}\lambda_{SH\Delta}^{2}+3\lambda_{H\Delta}^{2}
		\notag
		\\
		&+\lambda_{HS}^{2} + 3\lambda_{H\Delta}\lambda_{H\Delta}^{\prime}+\frac{5}{4}\lambda_{H\Delta}^{\prime 2}\notag\\
		&\Bigl. +\frac{3}{8}g_{1}^{4}+\frac{9}{8}g_{2}^{4} +\frac{3}{4}g_{1}^{2}g_{2}^{2}-(3g_{1}^{2}+9g_{2}^{2})\lambda_{H}
		\notag
		\\
		&-6y_{t}^{4}+12\lambda_{H}y_{t}^{2}\Bigr],\\
		%%%%%%%%%%%%%%%%%\lambda_{\Delta}
		\frac{\text{d}\lambda_{\Delta}}{\text{d}t}&=\frac{1}{16\pi^2}\Bigl[28\lambda_{\Delta}^{2}+24\lambda_{\Delta}\lambda_{\Delta}^{\prime}+6\lambda_{\Delta}^{\prime 2}+2\lambda_{H\Delta}^{2}
		\notag
		\\
		&+2\lambda_{H\Delta}\lambda_{H\Delta}^{\prime}+\lambda_{S\Delta}^{2}+6g_{1}^{4}+15g_{2}^{4}\Bigr.\notag\\
		&\left.-12g_{1}^{2}g_{2}^{2}-\left(12g_{1}^{2}+24g_{2}^{2}\right)\lambda_{\Delta}\right],\\
		%%%%%%%%%%%%%%%%\lambda_{\Delta}^{\prime}
		\frac{\text{d}\lambda_{\Delta}^{\prime}}{\text{d}t}&=\frac{1}{16\pi^2}\Bigl[18\lambda_{\Delta}^{\prime 2}+24\lambda_{\Delta}\lambda_{\Delta}^{\prime}+\lambda_{H\Delta}^{\prime 2}-6g_{2}^{4}
		\notag
		\\
		&+24g_{1}^{2}g_{2}^{2}-\left(12g_{1}^{2}+24g_{2}^{2}\right)\lambda_{\Delta}^{\prime}\Bigr],\\
		%%%%%%%%%%%%%%%%%\lambda_{S}
		\frac{\text{d}\lambda_{S}}{\text{d}t}&=\frac{1}{16\pi^2}\Bigl[20\lambda_{S}^{2}+2\lambda_{HS}^{2}+3\lambda_{S\Delta}^{2}\Bigr],\\
		%%%%%%%%%%%%%%%\lambda_{H\Delta}
		\frac{\text{d}\lambda_{H\Delta}}{\text{d}t}&=\frac{1}{16\pi^2}\Biggl[3g_{1}^{4}+6g_{2}^{4}-6g_{1}^{2}g_{2}^{2} +6\lambda_{H\Delta}y_{t}^{2}
		\notag
		\\
		&-\left(\frac{15}{2}g_{1}^{2}+\frac{33}{2}g_{2}^{2}\right)\lambda_{H\Delta}+12\lambda_{H}\lambda_{H\Delta}
		\notag
		\\
		&+4\lambda_{H}\lambda_{H\Delta}^{\prime} +4\lambda_{H\Delta}^{2}+16\lambda_{\Delta}\lambda_{H\Delta}
		\notag
		\\
		&+12\lambda_{\Delta}^{\prime}\lambda_{H\Delta}
+\lambda_{H\Delta}^{\prime 2}+6\lambda_{\Delta}\lambda_{H\Delta}^{\prime}
		\notag
		\\
		&\Bigl.+2\lambda_{\Delta}^{\prime}\lambda_{H\Delta}^{\prime}+2\lambda_{HS}\lambda_{S\Delta} \Biggr],
		\\
		%%%%%%%%%%%%%%%%\lambda_{H\Delta}^{\prime}
		\frac{\text{d}\lambda_{H\Delta}^{\prime}}{\text{d}t}&=\frac{1}{16\pi^{2}}\Bigr[12g_{1}^{2}g_{2}^{2}-\left(\frac{15}{2}g_{1}^{2}+\frac{33}{2}g_{2}^{2}\right)\lambda_{H\Delta}^{\prime}
		\notag
		\\
		&+4\lambda_{H}\lambda_{H\Delta}^{\prime}+8\lambda_{H\Delta}\lambda_{H\Delta}^{\prime}+4\lambda_{H\Delta}^{\prime 2}
		\notag
		\\
		&+4\lambda_{\Delta}\lambda_{H\Delta}^{\prime}+8\lambda_{\Delta}^{\prime}\lambda_{H\Delta}^{\prime}+2\lambda_{SH\Delta}^{2}
		\notag
		\\
		&+6\lambda_{H\Delta}^{\prime}y_{t}^{2}\Bigr],\\
		%%%%%%%%%%%%%%%%\lambda_{HS}
		\frac{\text{d}\lambda_{HS}}{\text{d}t}&=\frac{1}{16\pi^2}\Bigl[4\lambda_{HS}^2+8\lambda_{HS}\lambda_{S} + 12\lambda_{H}\lambda_{HS}
		\notag
		\\
		&+6\lambda_{S\Delta}\lambda_{H\Delta}+3\lambda_{S\Delta}\lambda_{H\Delta}^{\prime}+3\lambda_{SH\Delta}^{2}
		\notag
		\\
		&
		-\left(\frac{3}{2}g_{1}^{2}+\frac{9}{2}g_{2}^{2}\right)\lambda_{HS}+6\lambda_{HS}y_{t}^{2}\Bigr],
		\\
		%%%%%%%%%%%%%%%%%%%\lambda_{S\Delta}
		\frac{\text{d}\lambda_{S\Delta}}{\text{d}t}&=\frac{1}{16\pi^2}\Bigl[4\lambda_{S\Delta}^2+\lambda_{HS}(4\lambda_{H\Delta}+2\lambda_{H\Delta}^{\prime})
		\notag
		\\
		&+\lambda_{S\Delta}(16\lambda_{\Delta}+12\lambda_{\Delta}^{\prime}+8\lambda_{S})+\lambda_{SH\Delta}^{2}\Bigr.
		\notag
		\\
		&\Bigl.-(6g_{1}^{2}+12g_{2}^{2})\lambda_{S\Delta}\Bigr],\\ 
		%%%%%%%%%%%%%%%%%%%\lambda_{SH\Delta}
		\frac{\text{d}\lambda_{SH\Delta}}{\text{d}t}&=\frac{1}{16\pi^2}\Bigl[4\lambda_{H}+4\lambda_{H\Delta}+6\lambda_{H\Delta}^{\prime}
		\notag
		\\
		&+4\lambda_{HS} +2\lambda_{S\Delta}+6y_{t}^{2}-\frac{9}{2}g_{1}^{2}
		\notag
		\\
		&-\frac{21}{2}g_{2}^{2}\Bigr]\lambda_{SH\Delta},
\end{align}
where $g_{1}$, $g_{2}$, $g_{3}$ are the gauge coupling of $U(1)_{Y}$, $SU(2)_{L}$, and $SU(3)_{c}$, respectively.

%%%%%%%%%%%%%%%%%%%%%%%%%%%%%%%%%%%

\end{appendices}

\newpage

%%%%%%%%%%%%%%%%%%%%%%%%%%%%%%%%%%%%
\bibliography{siis}
\end{document}